\documentstyle[twoside,fleqn,epsf,espcrc2]{article}

\def\bit{\begin{itemize}}
\def\eit{\end{itemize}}

 \def\ss{\sigma}
 \def\ll{\lambda}

 \def\ll{\lambda}

 \def\gg{\gamma}
 \def\kk{\kappa}

 \def\bb{\beta}

 \def\beq{\begin{equation}}
 \def\eeq{\end{equation}}
 
 \def\fr{\frac}
 \def\Tr{{\mbox{Tr}}}

\newcommand{\AmS}{{\protect\the\textfont2
  A\kern-.1667em\lower.5ex\hbox{M}\kern-.125emS}}

\begin{document}
\title{Strong Coupling Model for String Breaking on the Lattice.}
\author{I T Drummond,  
        Department of Applied Mathematics and Theoretical Physics, \\
        University of Cambridge,
        Silver Street,
        Cambridge, England CB3 9EW}

\begin{abstract}
A model for $SU(n)$ string breaking on the lattice is formulated using
strong coupling ideas. It gives an explicit picture of string breaking, 
in the presence of dynamical quarks, as a mixing process between a string 
state and a two-meson state.  An analysis of the Wilson loop shows that
the evolution of the mixing angle as a function of separation may
obscure the expected crossover effect. If a sufficiently extensive mixing
region exists then an appropriate combination of transition amplitudes may help
to reveal the effect.  The sensitivity of the mixing region to the values 
of the meson energy and the dynamical quark mass is explored. 

\end{abstract}

\maketitle

\section{Introduction}
Much effort is being devoted to dynamical quarks \cite{DQ1,DQ2,DQ3,DQ4,DQ5,DQ6}.
For a review see S G\"usken \cite{DQ7}. For currently feasible values of the quark mass 
it is not expected that there will be dramatic effects on the computed hadron spectrum.
String breaking is an effect that is {\it only} possible in the presence of dynamical quarks.
Here we show explicitly that string breaking does occur on the lattice
and shows up as a mixing phenomenon.

The physical situation envisaged in lattice calculations comprises
a static quark and anti-quark separated by a spatial distance of $R$
lattice units. The two
static particles may either support a gauge string that runs between
them or separately bind a quark and an anti-quark to create a two-meson state.
When the string is stretched sufficiently, its energy coincides with
the energy of the two meson state. In the neighbourhood of this
critical separation the two different physical states mix permitting
the occurence of transitions between them.

\section{Strong Coupling Model}

We recall the rules for evaluating
simple graphs in the strong coupling limit of $SU(n)$ gauge theory \cite{CREUTZ,MonMun}.
\bit
\item[1.] A factor of $ \left(\bb/2n^2\right) $ for each plaquette.
\item[2.] A factor of $ 2\kk\left(1+\gg.e\right)/2 $
for each Wilson quark line in the direction of the unit vector $e$~.
Here $\kk$ is the standard quark hopping parameter.
\item[3.] Factors of $(-1)$ and $1/n$ and a trace over the spin matrix factors for 
each internal quark loop.
\eit
\subsection{Simple String Model}

In the absence of dynamical quarks the correlation function
of a string of length $R$ over an imaginary time interval $T$ is the standard
$R\times T$ Wilson loop. In leading strong coupling approximation the above
rules give for this string-string
propagator
\beq
{\cal G}_{SS}(T)=\left[e^{-\ss R}\right]^T~~,
\eeq
where the dimensionless string tension $\ss=-\log\left(\bb/2n^2\right)$~. 
The energy of the string state is $V(R)=\ss R$~.

\subsection{Model for Mesons}

Our model for mesons is one in which a light quark propagates 
along a static (anti-)quark line with a hopping parameter $\kk'$
that encodes the energy of the static bound state. 
Using this rule, we find that the meson propagator
has the structure $ g(T)=(1+\gg_0)/2\left(2\kk'\right)^T~.$
In other parts of the diagrams of the model the hopping parameter retains 
the value $\kk$ related to the light quark mass.

The propagator for two mesons moving independently, each bound to
its static quark, is
\newline\newline
\leftline{$g^{(1)}(T)\otimes g^{(2)}(T)$}
\beq
       = \left(\fr{1+\gg_0}{2}\right)^{(1)}\otimes
        \left(\fr{1-\gg_0}{2}\right)^{(2)}\left((2\kk')^2\right)^T~~.
\eeq

In fact only a particular combination of quark-anti-quark spin wave 
functions is relevant to the mixing phenomenon.
It is obtained by completing the light quark loop with the matrices
$(1\pm\gg_1)/2~,$ and including a factor of $(-1)$~. The two-meson 
propagator becomes
\beq
{\cal G}_{MM}(T)=\fr{1}{2}\left((2\kk')^2\right)^T~~.
\label{MESONS}
\eeq
We have then $ (2\kk')^2=e^{-E_M}~, $
where $E_M$ is the combined energy of the two bound mesons.

\section{String Breaking}

\begin{figure}[htb]
\begin{center}\leavevmode
\epsfxsize=69 truemm\epsfbox{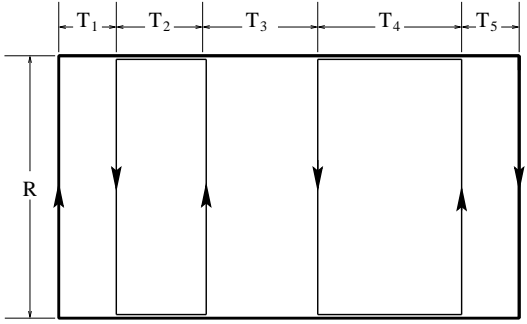}
\end{center}
\caption[]{Wilson loop (heavy line) containing internal quark loops (light lines).}
\label{figure:F2}
\end{figure}

Our model for string breaking involves summing over planar graphs
that incorporate transitions between the string and meson states.
A typical graph is shown in Fig 1. The corresponding contribution to the Green's
function is
$$
\left[e^{-\ss R}\right]^{T_1}\left[\left(2\kk'\right)^2\right]^{T_2}
   \left[e^{-\ss R}\right]^{T_3}\left[\left(2\kk'\right)^2\right]^{T_4}
     \left[e^{-\ss R}\right]^{T_1}
$$
$$
 \left(-\fr{t}{n}\right)\left[\left(2\kk\right)^2\right]^R[\sqrt{W}]^4
 \left(-\fr{t}{n}\right)\left[\left(2\kk\right)^2\right]^R[\sqrt{W}]^4~~,
$$
where
$$
t=\Tr\left(\fr{1+\gg_0}{2}\right)\left(\fr{1+\gg_1}{2}\right)
     \left(\fr{1-\gg_0}{2}\right)\left(\fr{1-\gg_1}{2}\right)~.
$$
The factors involving $\kk$ (as opposed to $\kk'$) are associated with the
{\it vertical} sides of the quark loops and the factor $\sqrt{W}$, where $W$ is an
energy, measures the rate of transition from a meson state to $q\bar{Q}$ pair. 
(This factor was omitted in the original formulation \cite{ITD}.) Note that the trace 
of the internal quark loop, which has the value $t=-\fr{1}{2}$, is equivalent 
to the quark-anti-quark spin projection mentioned previously. 

The structure of our diagrams is as follows:

\bit
\item[1.] A factor $a=e^{-\ss R} $ that
propagates the string by one time step.
\item[2.] A factor $b=\left(2\kk'\right)^2 $ that
propagates the two-meson state by one time step.
\item[3.] A factor $c=(W/\sqrt{2n})\left(2\kk\right)^R $
associated with the transition from string to two-meson state
and vice-versa.
\eit

To describe the transition from an initial time
to time $T$ we need a $2\times 2$ matrix of transition amplitudes
\beq
G(T)=\left(\begin{array}{cc}G_{SS}(T)&G_{SM}(T)\\ G_{MS}(T)&G_{MM}(T)\end{array}\right)~.
\eeq
then the above stepping procedure can be represented by
$
G(T+1)=AG(T)~~,
$
where the matrix $A$ is given by
\beq
A=\left(\begin{array}{cc}a&ac\\ bc&b\end{array}\right)~~.
\eeq
If for definiteness  we set $ G(0)=1$, then $ G(T)=\left(A\right)^T$~.
We can set $A=DO\Lambda O^{-1}D^{-1}$ where 
\beq
D=\left(\begin{array}{cc}\sqrt{a}&0\\0&\sqrt{b}\end{array}\right)~,
O=\left(\begin{array}{cc}\cos\theta&-\sin\theta\\\sin\theta&\cos\theta\end{array}\right)~,
\eeq
\beq
\Lambda=\left(\begin{array}{cc}\ll_{+}&0\\0&\ll_{-}\end{array}\right)~~.
\eeq
The entries in $\Lambda$ are the eigenvalues of $A$
\beq
\ll_{\pm}=\fr{1}{2}\left\{(a+b)\pm\sqrt{(a-b)^2+4abc^2}\right\}~~.
\eeq
The corresponding eigenvectors are the columns of $DO$. The mixing
angle $\theta$ is
\beq
\tan\theta=\fr{-(a-b)+\sqrt{(a-b)^2+4abc^2}}{2\sqrt{ab}~c}~~.
\eeq

\section{Transition Amplitudes}

We assume that each transition amplitude begins and ends with an
appropriate propagator. We find for the Wilson loop
\beq
{\cal G}_{SS}(T)=(1,0)A^{T-1}\left(\begin{array}{c}a\\0\end{array}\right)
\eeq
with the result
\beq
{\cal G}_{SS}(T)=a\left(\cos^2\theta~\ll_{+}^{T-1}+\sin^2\theta~\ll_{-}^{T-1}\right)~~.
\eeq
Similar reasoning yields
\beq
{\cal G}_{MM}(T)=b\left(\sin^2\theta~\ll_{+}^{T-1}+\cos^2\theta~\ll_{-}^{T-1}\right)~~,
\eeq
\beq
{\cal G}_{MS}(T)=\sqrt{ab}\sin\theta\cos\theta\left(\ll_{+}^{T-1}-\ll_{-}^{T-1}\right)~~.
\eeq

All amplitudes see both exponentials, $ e^{-E_{\pm}T}=\ll_{\pm}^T~~.$
The the lower energy exponential should dominate the asymptotic behaviour.
However what is observed will be influenced by the behaviour of the mixing angle.
When $\ss R<<E_M$ then $a>>b$ and $\theta\simeq 0$~,$E_{+}\simeq \ss R$~,$E_{-}\simeq E_M$~.
Under these circumstances the coupling to the state with energy $E_{-}$ will vanish 
and only the exponential associated with $E_{+}$ will be observed as expected. 
When $\ss R>>E_M$ then $b>>a$ and $\theta\simeq \fr{\pi}{2}$~,$E_{+}\simeq E_M$~,
$E_{-}\simeq \ss R$~. Now the
coupling to the state with energy $E_{+}$ will vanish and only the exponential
associated with $E_{-}$ will be observed. That is, both above and below the crossover
point, only the {\it original} string behaviour $ e^{-V(R)T}~, $
will be observed. The movement of the mixing angle therefore obscures the crossover phenomenon
in which the string energy is expected to be bounded by the energy of the
two-meson state. Complementary results hold for ${\cal G}_{MM}(T)$~. It is clear that
${\cal G}_{SM}(T)$ is suppressed {\it outside} the mixing region, where 
$\sin\theta\cos\theta\simeq 0$~.

\section{Mixing Region}

To see mixing effects directly in the Wilson loop the mixing
region in $R$, for which both $\sin\theta$ and $\cos \theta$ 
are reasonably large, must be of sufficient size to be resolved 
on the lattice. The range of the mixing region in $R$ can be estimated as
\beq
\Delta R=\left(\fr{\pi}{2}\fr{dR}{d\theta}\right)_{R=R_c}
=\pi\sqrt{\fr{2}{3}}\fr{W}{\ss}e^{-m_qE_M/\ss}~~.
\label{dR}
\eeq
If we estimate the parameters in the model as  $a^{-1}=1.5$ GeV, $\ss^{1/2}=0.427$ GeV,
$W\simeq E_M=1.0$ GeV then the critical distance is $R_c=8.22$ lattice spacings. For
$m_q$=.1 GeV we then find $\Delta R$=18.3 lattice spacings, while for $m_q$=.5 GeV we obtain
$\Delta R$=2.0 lattice spacings. We see then that if the dynamical quark is light enough
and the meson energy not too high there is a reasonable chance of observing
the string breaking within the mixing region. The exponential factor in eq(\ref{dR})
guarantees that if the meson energy is larger than 1 GeV the mixing region will be 
rapidly restricted as the value of $m_q$ is raised. An appropriate combination of 
transition amplitudes may also help in identifying the mixing process. For example
in the model the combination $b{\cal G}_{SS}(T)+a{\cal G}_{MM}(T)$ removes the
influence of the mixing angle variation.

\end{document}